\documentclass[twocolumn,aps,amsmath,amssymb]{revtex4-2}
\usepackage{mathtools, xcolor}
\usepackage{booktabs, siunitx}
\usepackage{appendix}
\usepackage[normalem]{ulem}
\usepackage[colorinlistoftodos]{todonotes}
\usepackage{soul}
\usepackage{makecell}
\usepackage{physics}
\usepackage{amsmath}
\usepackage{amsmath, amssymb}
\usepackage{tikz}
\usepackage{braket}
\usepackage[unicode=true, colorlinks=true, citecolor={blue!80!black}, urlcolor={blue!50!black}, linkcolor = {blue!80!black}]{hyperref}

\newcommand{\squaremarker}{\protect\tikz{\protect\draw (0,0) rectangle (0.15,0.15);}}
\newcommand{\circlemarker}{\protect\tikz{\protect\draw (0.075,0.075) circle (0.075);}}
\newcommand{\diamondmarker}{\protect\tikz{\protect\draw[rotate=45] (0,0) rectangle (0.15,0.15);}}
\begin{document}

\title{FerBo: a noise resilient qubit hybridizing Andreev and fluxonium states}

\newcommand{\affA}{\affiliation{Departamento de F\'{\i}sica Te\'orica de la Materia Condensada, \mbox{Condensed Matter Physics Center (IFIMAC)} and Instituto Nicol\'as Cabrera, Universidad Aut\'onoma de Madrid, 28049 Madrid, Spain}}
\newcommand{\affB}{\affiliation{Quantronics group, Service de Physique de l'\'Etat Condens\'e \mbox{(CNRS, UMR 3680)}, IRAMIS, CEA-Saclay, Universit\'e Paris-Saclay, 91191 Gif-sur-Yvette, France}}
\newcommand{\eqContrib}{\thanks{These authors contributed equally to this work.}}
\author{J. J. Caceres}
\eqContrib
\affB
\author{F. J. \surname{Matute-Ca\~nadas}}
\eqContrib
\affA
\author{D. Sanz Marco}
\affB
\author{J. Ortuzar}
\affB
\author{E. Flurin}
\affB
\author{C. Urbina}
\affB
\author{H. Pothier}
\affB
\author{A. \surname{Levy Yeyati}}
\affA
\author{M. F. Goffman}
\email[Corresponding author: ]{marcelo.goffman@cea.fr}
\affB
\date{\today}
\begin{abstract}

We propose a novel superconducting quantum circuit that should be robust against both relaxation and dephasing over a wide and experimentally accessible parameter range. The circuit consists of a parallel arrangement of a large inductance, a small capacitor, and a well-transmitting Josephson weak link. Protection against relaxation arises from the hybridization between the fermionic degree of freedom associated with Andreev levels in the weak link and the bosonic electromagnetic mode of the LC circuit, hence its name: FerBo. Furthermore, as in the fluxonium qubit, delocalization of the wavefunctions in phase space provides resilience against dephasing.
\end{abstract}
\maketitle

\section{Introduction}

Superconducting circuits are a leading platform for quantum information processing, offering scalability and versatility but suffering from pronounced sensitivity to environmental noise. This noise limits qubit coherence, and thus computation fidelity, motivating the development of noise mitigation strategies. Active error correction relying on repeated measurements and feedback can, in principle, correct errors but incurs significant overhead, requiring many physical qubits per each logical one \cite{riste2015detecting, leghtas2015confining}. A complementary approach is passive, hardware-level protection, which aims to encode quantum information in qubits that are by design less sensitive to noise \cite{douccot2012physical, brooks2013, Smith2020}.

\begin{figure}[h!]
    \centering
    \includegraphics[width=0.45\textwidth]{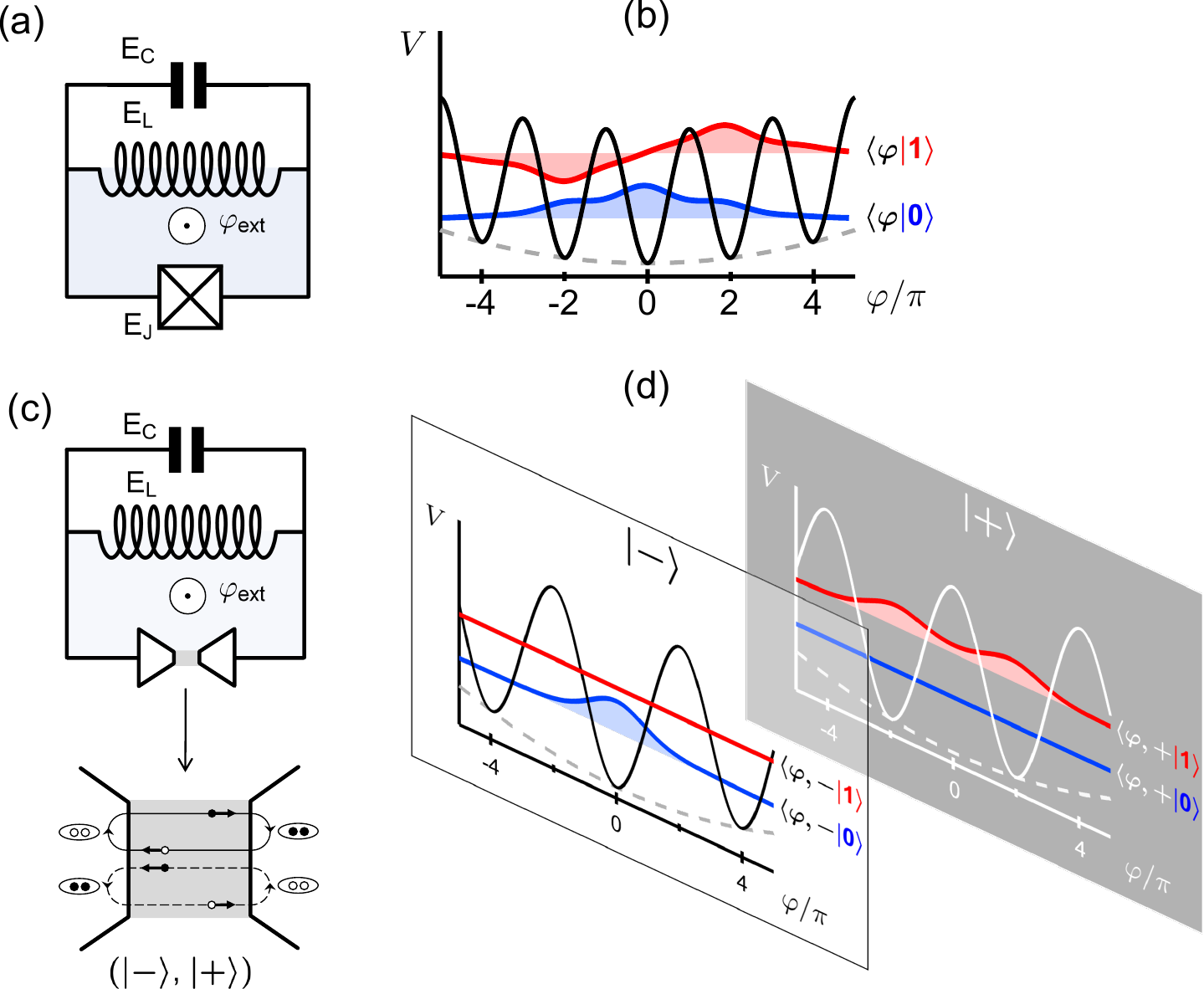}
    \caption{Principle of protection. (a) Electrical circuit for the light-fluxonium qubit. (b) Potential energy $V(\varphi)$ of a ``light'' fluxonium ($Z\gg R_Q$) for $\varphi_{\rm ext}=0$, together with the two lowest eigenstates wavefunctions. $V(\varphi)$ is the sum of a parabolic term associated to the inductance (dashed line) and a $\cos(\varphi)$ term associated to the Josephson energy. The delocalization of the wavefunctions in phase strongly reduces the qubit's sensitivity to external flux noise.  (c) Idealized FerBo circuit: light fluxonium incorporating an Andreev ballistic weak link. Andreev reflections at the normal-superconducting interfaces give rise to the formation of Andreev bound states ($\ket{-},\ket{+}$). (d) FerBo  potential depends on Andreev level occupation. The Josephson cosine term is replaced by $\mp\cos(\varphi/2)$ terms associated with the two Andreev bound states. The FerBo ground-state wavefunction resides in the $\ket{-}$ Andreev manifold, whereas its first excited state resides in the $\ket{+}$ one. Their simultaneous phase delocalization and disjoint support in the Andreev sector suppress both flux-noise-induced dephasing and charge-noise-induced relaxation.}
   \label{fig:protection}
\end{figure}

Such protection strategies have been explored in the most common family of superconducting qubits, where information is stored in collective electromagnetic modes of Josephson circuits built from capacitors, inductors, and tunnel junctions. Notable examples include the transmon \cite{Koch2007} and fluxonium \cite{Manucharyan2009} qubits. The transmon, a Cooper pair box operated in a regime where its wavefunctions are delocalized in charge, is exponentially less sensitive to charge noise dephasing. Conversely, the fluxonium, a flux qubit with extended phase wavefunctions, achieves suppression of flux noise dephasing sensitivity. However, in both cases, delocalization increases transition matrix elements between qubit states. While dephasing is suppressed, the coupling to dissipative channels is enhanced, ultimately limiting the overall coherence time \cite{Gyenis_2021}.

Simultaneous protection against both dephasing and relaxation can be achieved in circuits with multiple degrees of freedom, where the qubit wavefunctions exhibit both disjoint support and reduced sensitivity to external parameters. The $0-\pi$ qubit \cite{brooks2013,Groszkowski_2018}, the $\cos(2\varphi)$ qubit \cite{Smith2020,Smith2022}, the GKP qubit \cite{Sellem2025} or the bifluxon qubit \cite{Kalashnikov2020}, among others \cite{Gyenis_2021,Hays2025}, exemplify this principle by combining charge and flux degrees of freedom such that wavefunction delocalization along one coordinate suppresses dephasing, whereas localization along the other ensures suppressed relaxation. While this approach demonstrates the principle of dual protection, it demands stringent parameter choices, making experimental realization challenging \cite{gyenis_2021experimental}.

Here, we propose an alternative strategy in which the Josephson tunnel junction of the fluxonium is replaced by a superconducting weak link hosting Andreev bound states (ABS). As we explain below, exploiting the additional fermionic degree of freedom associated with the even occupation of ABS \cite{Desposito2001,Zazunov2003,Janvier2015,Hays2018} in the FerBo enables simultaneous protection against both relaxation and dephasing. 
The protection principle is illustrated in Fig.~\ref{fig:protection}. Panel (a) shows the conventional fluxonium circuit. The Hamiltonian of the circuit reads
\begin{equation}{\label{eq:hamiltonian_fluxonium}}
    \hat H = 4 E_C \hat{n}^2 + \frac{1}{2} E_L (\hat{\varphi}- \varphi_{\text{ext}})^2 - E_J\; \mathrm{cos}(\hat{\varphi}),
\end{equation}
where $\hat{\varphi}$ is the phase operator and $\hat{n}$ the conjugate Cooper pair charge number operator. $E_C$, $E_L$, and $E_J$ are the capacitive, inductive and the Josephson energies of the circuit, respectively.  Panel (b) shows the associated potential energy $V(\varphi)$ resulting from the sum of the last two terms in the Hamiltonian: the parabolic term associated with $E_L$ and the cosine term related to the Josephson tunnel junction. We focus on the ``light fluxonium'' regime $E_L\ll E_J,E_C$ and at $\varphi_{\rm ext}=0$ \cite{mencia2024integerfluxoniumqubit}. In the same panel, the two lowest eigenstates delocalized over multiple wells are depicted. Delocalization of the eigenstates suppresses  dephasing due to flux noise. 

In our proposed FerBo circuit, we consider first a ballistic single-channel weak link hosting Andreev states $(\ket{-},\ket{+})$ (panel (c)) with energies proportional to $\mp \cos(\hat{\varphi}/2),$ respectively. The  hybridization with the LC mode produces Andreev-state-dependent potentials associated with the $\ket{-}$ and $\ket{+}$ manifolds, in which the locations of minima and maxima are inverted  (panel (d)). Let us call $\ket{g_-}$ and $\ket{g_+}$ the ground states in the potentials associated with $\ket{-}$ and $\ket{+}$, and $\ket{e_{-,+}}$ their respective first excited states. In the ``light'' regime in which the wavefunctions are delocalized over several wells, the global ground state $\ket{0}$ is $\ket{g_-}$ and the first excited state $\ket{1}$ is $\ket{g_+}$. Because their wavefunctions reside in different Andreev sectors, $\ket{0}$ and $\ket{1}$ have disjoint supports, which provides protection against relaxation induced by flux and charge noise. Protection against dephasing results from the delocalization of the wavefunctions in phase space. In the case of a non-ballistic weak link, the two Andreev sectors mix. However, as explained below, protection against relaxation persists as long as $\ket{0}$ and $\ket{1}$ retain the same phase parity.

The manuscript is organized as follows. Section~\ref{sec:model} introduces the theoretical model of the FerBo qubit based on the microscopic theory of ABS. Section~\ref{sec:protection} presents analytical and numerical results for the qubit's energy spectrum. It provides a systematic analysis of the relaxation  and dephasing susceptibilities of the FerBo circuit to evaluate its hardware-level noise immunity and identify optimal design parameters.  Section~\ref{sec:exp} discusses experimental considerations and possible implementations. Finally, Section~\ref{sec:conclusions} concludes with an outlook.

\section{Model}\label{sec:model}

In a tunnel-barrier weak link, the ABS energies lie very close to the superconducting gap edge so that the circuit lowest energy states lie in the same Andreev sector. In contrast, in quasi-ballistic extended junctions the ABS lie well within the superconducting gap, and their dynamics play a major role \cite{Hays2018} in the electrodynamic response of the Josephson device. Following Refs.~\cite{Bretheau2014,Vakhtel2024,gungordu2025},  we describe the circuit with an effective Hamiltonian that couples the bosonic circuit degree of freedom $(\hat \varphi,\hat n)$ with the Andreev levels:

\begin{equation}{\label{eq:hamiltonian}}
   \hat H = 4 E_C \hat{n}^2 + \frac{1}{2} E_L \hat{\varphi}^2 + H_{\rm WL}(\hat{\varphi} - \varphi_{\text{ext}}),
\end{equation}
 where the term associated with the weak link $H_{\rm WL}$ describes the Andreev element as a single spin-degenerate energy level coupled to two superconducting leads. The parameters of this Josephson quasi-resonant level (JQRL) junction are the coupling to the superconducting electrodes $\Gamma_L$ and $\Gamma_R$, the energy of the level $\epsilon_r$, measured with respect to the Fermi level of the electrodes, $\Gamma = \Gamma_L + \Gamma_R$ the effective coupling of the single level dot to the leads, and $\delta\Gamma = \Gamma_L - \Gamma_R$ the coupling asymmetry. When $\Gamma_L$ and $\Gamma_R$ are much smaller than the superconducting gap of the leads $\Delta$, the term $H_{\rm WL}$ can be described with the following atomic limit potential in the even-parity Andreev manifold~\cite{kurilovich2021,Vakhtel2024,gungordu2025}:
\begin{equation}\label{eq:potential}
        H_{\rm WL}(\hat{\varphi})= \Gamma \cos(\hat{\varphi}/2) \hat \sigma_x -\delta \Gamma \sin(\hat{\varphi}/2) \hat \sigma_y+\epsilon_r \hat \sigma_z,
\end{equation}
where $\hat{\sigma}_{i}$ are the Pauli matrices acting on the even-parity subspace of the Andreev level. When $\varepsilon_r = \delta \Gamma = 0$, $H_{\rm WL}=\Gamma \cos(\hat \varphi/2) \hat \sigma_x$ and we recover the ballistic situation of Fig.~\ref{fig:protection}(c).

Note that we have omitted the effects of Coulomb interaction and internal dynamics of the level due to the capacitance between leads~\cite{gungordu2025,Kurilovich0221}, which are expected to be small. Anticipating on the practical implementation of the circuit in which the dominant capacitance is not that of the junction, the external flux $\varphi_{\text{ext}}$ was placed in the Andreev potential term of the Hamiltonian~\cite{You}.
\begin{figure}[t]
    \centering
    \includegraphics[width=\linewidth]{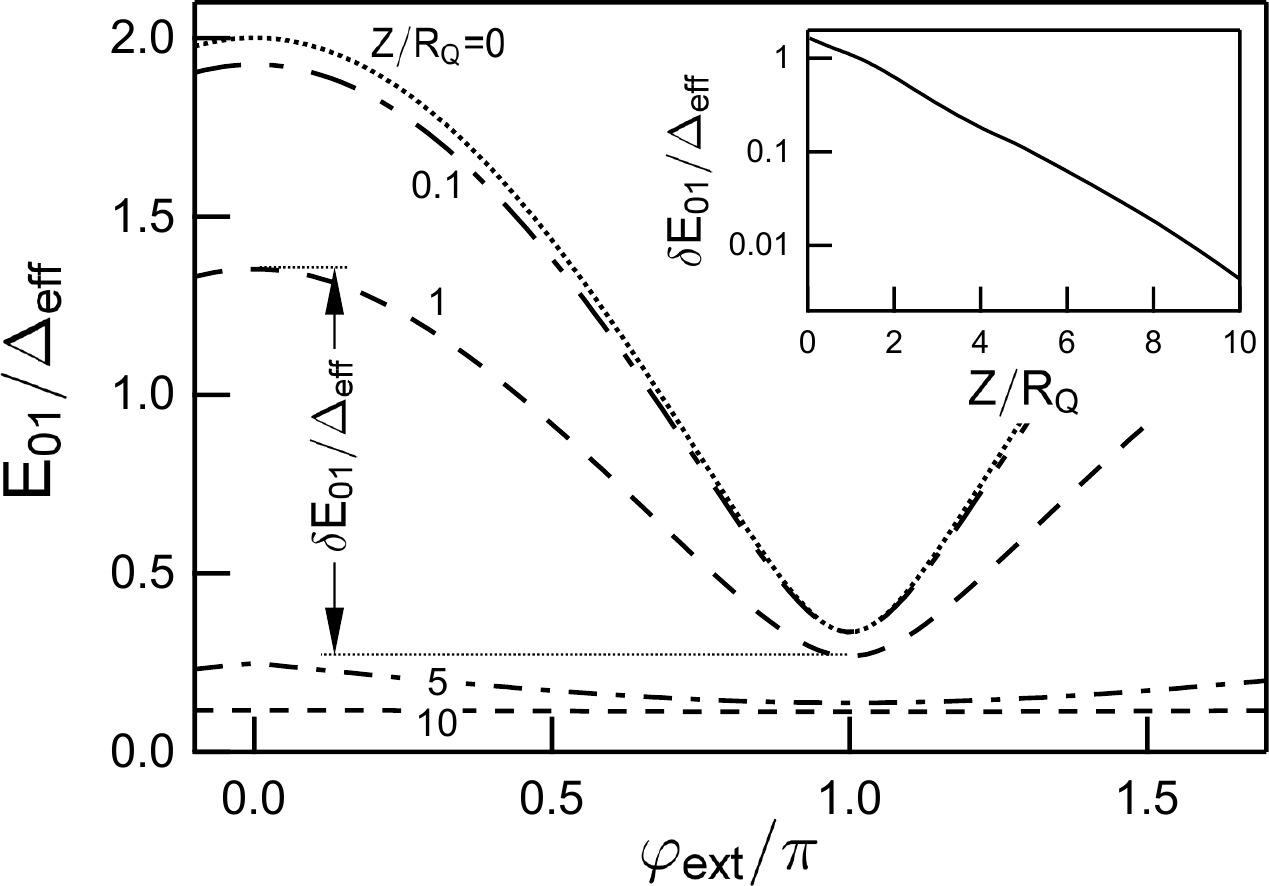}
    
    \caption{Transition energy $E_{01}$ between the ground and first excited states of the FerBo circuit normalized by $\Delta_{\mathrm{eff}}$ for $\Gamma/E_C=0.75$; $\epsilon_r/E_C=0.13$, $E_C=15$~GHz and $\delta\Gamma/\Gamma=0.01$) as a function of normalized external magnetic flux shown for different values of the $LC$-mode impedance $Z/R_Q$. The dotted line corresponds to the Andreev transition energy for perfect phase bias. Inset: flux dispersion amplitude $\delta E_{01}=E_{01}(0)-E_{01}(\pi)$  as a function of reduced impedance $Z/R_Q$.}
        \label{fig:E01_JRL-fluxonium}
\end{figure}

This effective Hamiltonian captures the relevant low-energy Andreev degree of freedom and allows a transparent analysis of the protection mechanism. Although simplified, this description is corroborated by a more detailed microscopic model of the weak link described in Appendix~\ref{app:tightbinding}. The microscopic calculation qualitatively identifies the same parameter regime in which protection occurs, thereby supporting the use of the effective model throughout the main text.
\begin{figure*}[t]
    \centering
    \includegraphics[width=\linewidth]{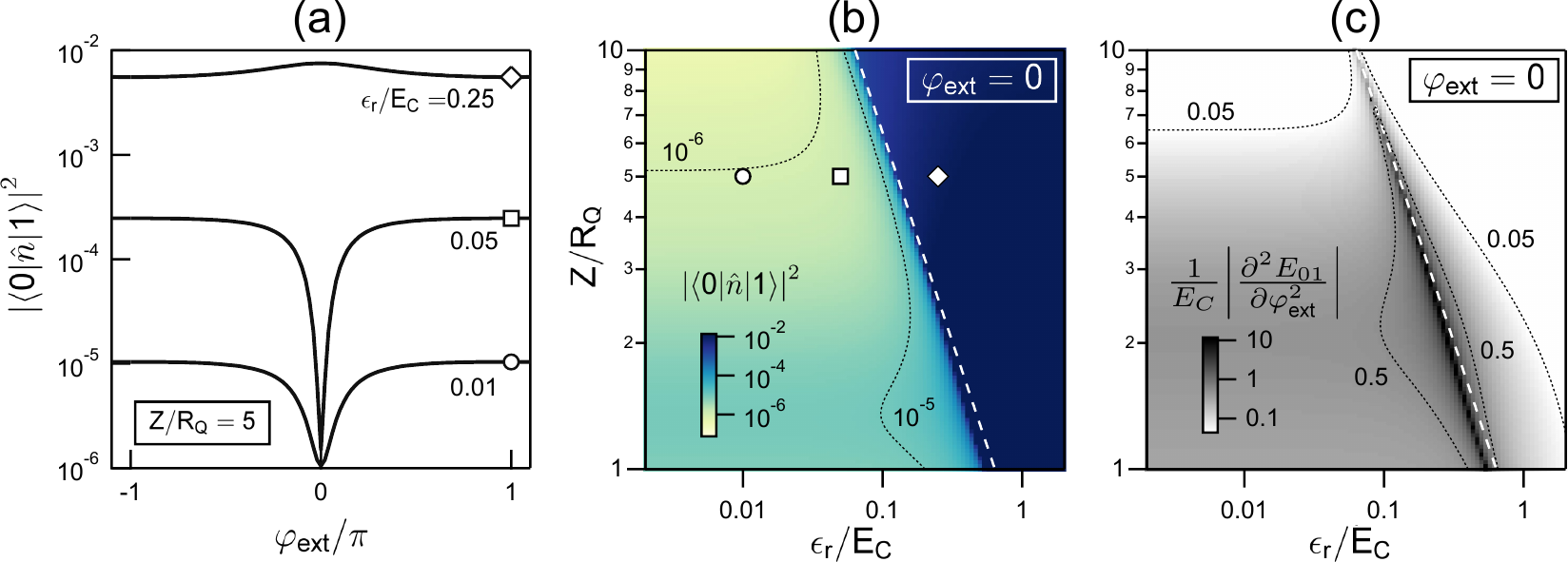}
    \caption{Relaxation and dephasing susceptibilities. (a) Normalized charge relaxation susceptibility $|\bra{0}\hat{n}\ket{1}|^2$ as a function of the external flux for three values of $\epsilon_r/E_C=0.01(\circlemarker), 0.05 (\squaremarker), 0.25(\diamondmarker)$ at fixed LC-impedance ($Z/R_Q=5$), $\delta\Gamma/\Gamma=0.01$ and $\Gamma/E_C=0.75$. At $\varphi_{ext}=0$ the susceptibility  drops sharply, and the circuit becomes protected against relaxation due to charge and flux noise (see main text). 
    (b) Color map of $|\bra{0}\hat{n}\ket{1}|^2$ at $\varphi_{\rm ext}=0$ as a function of level position $\epsilon_r/E_C$ and of the reduced impedance $Z/R_Q$. Light-colored regions indicate negligible susceptibility, whereas dark regions are unprotected. 
    White dashed line corresponds to $Z/R_Q=2 E_C/\pi \epsilon_r$. 
    (c) Grayscale map of the normalized second-order flux dephasing susceptibility $(1/E_C)|\partial^2 E_{01}/\partial \varphi_{\rm ext}^2|$. The FerBo regime corresponds to the upper-left region of the color plots, where the circuit is simultaneously protected against relaxation and dephasing. 
        }
        \label{fig:spectrum_vs_flux_diff_Z}
\end{figure*}

The energy scales $E_L$ and $E_C$  define the impedance of the $LC$ mode as $Z=(R_Q/\pi)\sqrt{2E_C/E_L},$ with $R_Q=h/4e^2$ the resistance quantum. In the small impedance limit ($Z\to 0$ reached with $L\to 0$), zero point phase fluctuations are negligible and the eigenvalues of the circuit are those of the weak link Hamiltonian $H_{\rm WL}(\varphi_{\rm ext})$, reproducing the standard expression for the energy of the Andreev bound states in a phase-biased Josephson junction, $\pm E_A$, and the associated pair transition between the ground $|+\rangle$ to the excited $|-\rangle$ Andreev manifolds: 
\begin{equation}
E_{01}(Z=0)=2E_A = 2\Delta_{\rm eff} \sqrt{1- \tau \sin^2(\varphi_{\rm ext} /2)}
\end{equation}
\noindent with the effective gap energy 
\begin{equation}
\Delta_{\rm eff} \equiv  \sqrt{\Gamma^2+\epsilon_r^2}
\label{eq:Delta_eff}
\end{equation}
and the junction transmission
\begin{equation}\label{eq:tau}
\tau \equiv \frac{\Gamma^2 - \delta\Gamma^2}{\Gamma^2 + \varepsilon_r^2}
=1-\frac{\epsilon_r^2+\delta\Gamma^2}{\Delta^2_{\rm eff}}.
\end{equation}

Figure~\ref{fig:E01_JRL-fluxonium} shows how the $\varphi_{\rm ext}$-dependence of the transition energy between the ground and the first excited state, obtained by numerical diagonalization of the Hamiltonian of Eq.~(\ref{eq:hamiltonian}) \cite{Caceres2025}, varies with the impedance $Z/R_Q$. As for the conventional fluxonium, the eigenenergies flux dispersion, and consequently that of the transition energies, decrease exponentially when increasing $Z$ (see inset of the same figure), which defines the ``light'' regime $Z\gg R_Q$. This has a straightforward impact on the coherence properties, as discussed in the next section.

\section{Hardware level protection against noise}\label{sec:protection}

Following the framework established for hardware-level noise protection \cite{Gyenis_2021}, two quantities characterize the qubit vulnerability to environmental noise: the susceptibility to relaxation and the susceptibility to dephasing. In fluxonium-like circuits, the dominant energy relaxation mechanism is coupling to two-level system defects in the capacitor dielectrics \cite{Nguyen2019,Ardati2024,Ateshian2025,Azar2026}. The susceptibility to relaxation is therefore quantified by the  matrix element $|\bra{0}(\partial \hat H/\partial \hat{n})\ket{1}|^2=|8E_C\bra{0}\hat{n}\ket{1}|^2$, which mediates the coupling to charge noise \footnote{In the FerBo, flux noise susceptibility is given by $|\bra{0}\partial \hat H_{\rm WL}/\partial \varphi_{\rm ext}\ket{1}|^2$. We find that its contribution to relaxation is negligible in the protected regime defined in section \ref{sec:protection}}. 

The susceptibility to dephasing is mainly associated with the flux dispersion of the qubit transition energy. At a first-order sweet spot where $\partial E_{01}/\partial \varphi_{\text{ext}}=0$, the leading contribution to dephasing is governed by the curvature $\partial^2 E_{01}/\partial \varphi_{\text{ext}}^2$. Minimizing this second derivative, ideally achieving exponentially flat dispersion, ensures that flux fluctuations induce minimal phase accumulation between the qubit states.

These two susceptibilities serve as the fundamental figures of merit for evaluating the hardware-level noise immunity of the FerBo circuit. By analyzing these quantities across the parameter space, we identify optimal design regimes that simultaneously suppress the qubit's coupling to both charge and flux noise channels. The results we present in the following are representative of the behavior of the circuit when $E_C\gtrsim\Gamma$ (see Appendix~\ref{sec:Gamma&deltaGamma}). We set the asymmetry of the Andreev weak link to $\delta\Gamma/\Gamma=0.01$ and vary $\epsilon_r$ to tune the transmission properties. This choice is motivated by experimental considerations: $\epsilon_r$ is directly linked to the chemical potential of the junction, which is the primary parameter controlled by electrostatic gating in most experimental implementations (see section~\ref{sec:exp}).

Figure~\ref{fig:spectrum_vs_flux_diff_Z}(a) shows the normalized charge relaxation susceptibility $|\langle0|\hat{n}\ket{1}|^2$ for three different values of $\epsilon_r$ as a function of the external flux in the high impedance regime ($Z/R_Q=5$). The results highlight the external flux to be chosen for relaxation protection: when $\epsilon_r$ is small enough, the susceptibility shows a sharp dip at $\varphi_\text{ext}=0$ and drops by four orders of magnitude. Figure~\ref{fig:spectrum_vs_flux_diff_Z}(b) shows the relaxation protection at zero external flux as a function of the main parameters of the circuit: the impedance $Z$ of the $LC-$mode and $\epsilon_r$, one of the parameters that controls the coupling between Andreev manifolds. The color plot clearly shows two distinct regions separated by a sharp boundary. When $E_C$ is the largest energy scale, the position of this boundary is well approximated by $\epsilon_r=\sqrt{2E_CE_L}$, or equivalently $Z/R_Q = 2E_C/(\pi\epsilon_r)$ (white dashed line in Fig.~\ref{fig:spectrum_vs_flux_diff_Z}, see also Appendix~\ref{sec:Gamma&deltaGamma}).
To the left of the boundary, charge relaxation susceptibility is low and the circuit is protected over a wide range of circuit parameters.  

As for the fluxonium, dephasing protection is achieved by increasing the impedance $Z$, which delocalizes the wavefunctions. This is illustrated in Fig.~\ref{fig:spectrum_vs_flux_diff_Z}(c), where the dephasing susceptibility $\partial^2 E_{01}/\partial \varphi_{\rm ext}^2$ is plotted as a function of $\epsilon_r/E_C$ and $Z/R_Q$. Increasing the impedance leads to a strong suppression of this second derivative, largely independent of $\epsilon_r$ (except close to the boundary line defined before). Consequently, a large transmission weak link embedded in a high-impedance circuit provides the best protection against both relaxation and pure dephasing.

Interestingly, Fig.~\ref{fig:spectrum_vs_flux_diff_Z}(b) shows that the protection against relaxation also exists for moderate values of $Z/R_Q$. The constraint of very small $\epsilon_r/E_C$ (high transparency) is then relaxed, but at the cost of a stronger dispersion of the states (Fig.~\ref{fig:E01_JRL-fluxonium}), which leads to a higher dephasing susceptibility (Fig.~\ref{fig:spectrum_vs_flux_diff_Z}(c)).


\begin{figure*}[t]
    \centering
    \includegraphics[width=\linewidth]{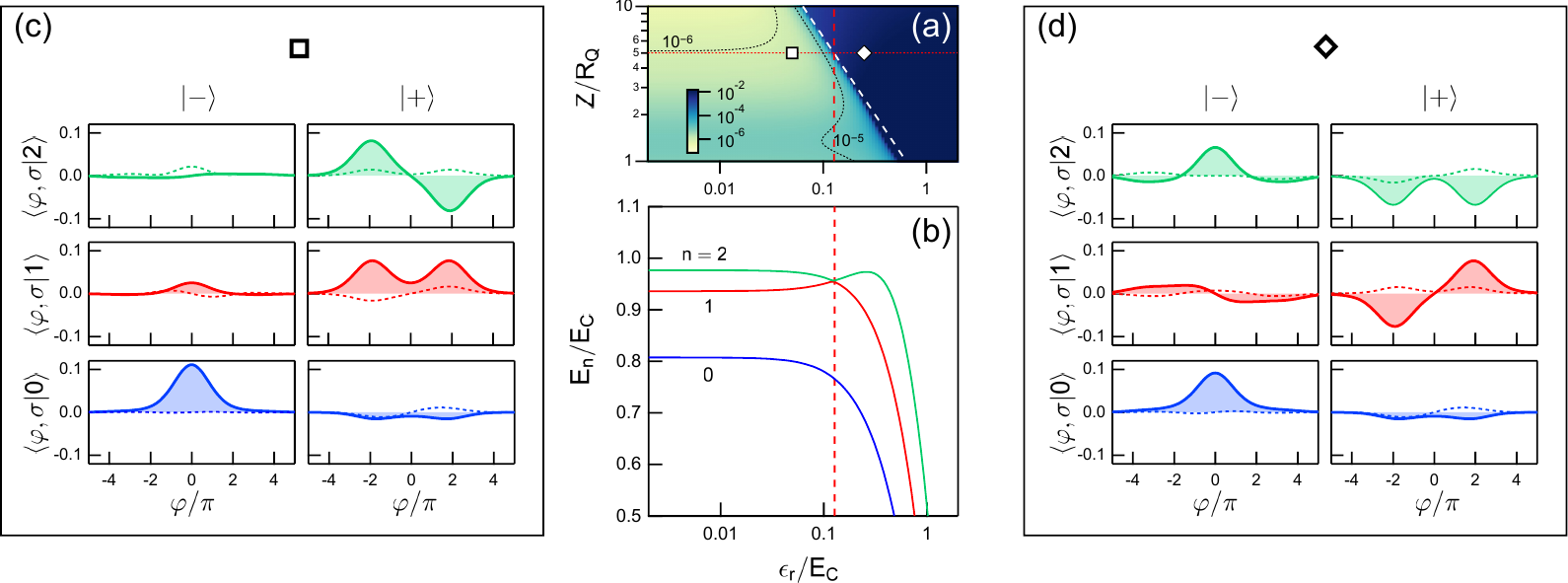}
    \caption{Symmetry transition underlying flux-noise protection. (a) Color map of the matrix element $\left|\langle \mathrm{0}|\hat{n}|\mathrm{1}\rangle\right|^{2}$as a function of the LC-impedance and the resonant level position $\epsilon_r$ (top) (b) Eigenenergies of the three lowest states of the FerBo ($\Gamma/E_C=0.75$ and $\delta\Gamma/\Gamma=0.01$ ) at a fixed impedance, $Z/R_Q=5$, depicted as a function of $\epsilon_r$. The sharp boundary between ``protected'' and ``unprotected'' regions coincides with the avoided crossing between the first and second excited states, highlighted by the red vertical dashed line. (c) and (d): eigenstates wavefunctions below and above the (avoided) crossing, respectively. Solid lines are the real parts; dashed lines (magnified by a factor of 20) are the imaginary parts. Wavefunctions are projected onto the $\ket{-}$ and $\ket{+}$ sectors. To the left of the anticrossing (c) the ground state (blue) is mainly localized  in the $\ket{-}$ sector and resembles $\ket{g_-}$, while the first excited state (red), with even symmetry, resides mostly in the $\ket{+}$ sector and resembles $\ket{g_+}$. The second excited state (green), with odd symmetry, is also predominantly located in the $\ket{+}$ sector and resembles $\ket{e_+}$. To the right of the  anticrossing (d), the symmetries swap: the first excited state becomes odd like $\ket{e_+}$, and the second excited state becomes even like $\ket{g_+}$. The same qualitative behavior occurs at fixed $\epsilon_r$ varying the LC-mode impedance Z.}
        \label{fig:ferbo_vs_phase}
\end{figure*}


The origin of the abrupt transition in relaxation susceptibility and the emergence of the protected region (Fig.~\ref{fig:spectrum_vs_flux_diff_Z}(b)) can be understood by examining the qubit wavefunctions in the ballistic basis $(\ket{-},\ket{+})$. Figure~\ref{fig:ferbo_vs_phase} displays the ground state and the first two excited states on both sides of the protection boundary, revealing a symmetry change that underlies the protection mechanism. Far inside the protected regime (Fig.~\ref{fig:ferbo_vs_phase}(c), open square in Fig.~\ref{fig:ferbo_vs_phase}(a), the ground state (blue) resembles $\ket{g_-}$ and has most of its weight in the $|-\rangle$ Andreev sector. The first (red) and second (green) excited states reside primarily in the $|+\rangle$ sector, and resemble $\ket{g_+}$ and  $\ket{e_+}$. This separation between Andreev manifolds provides protection against relaxation from any operator that does not couple the two sectors, i.e., from any purely bosonic operator.

In addition, loop-based quantum circuits exhibit some selection rules when biased at 0 or $\pi$ external flux. In our case, the first excited state has essentially the same parity as the ground state (even parity under phase inversion) with only a small imaginary part with opposite parity (dotted lines, magnified $\times 20$) arising from $\delta \Gamma \sin(\hat \varphi/2) \hat \sigma_y$. Wavefunctions with the same parity have suppressed matrix elements of operators that are odd under phase inversion, such as $\hat{\varphi}$ and $\hat n$, which constitute the dominant contributions to relaxation. Consequently, the residual overlap arising from the imperfect separation between Andreev sectors (which is unavoidable when $\epsilon_r \ne 0$) is further suppressed -- it would be completely suppressed for $\delta\Gamma=0$, see Appendix~\ref{sec:Gamma&deltaGamma}.

When $\epsilon_r$ increases, the term $\epsilon_r \hat \sigma_z$ in the Hamiltonian couples states living in opposite sectors. Due to the symmetries of the wavefunctions, $\ket{1}\approx\ket{g_+}$ is pushed up in energy by $\ket{0}\approx\ket{g_-}$, while $\ket{2}\approx\ket{e_+}$ is pushed down by $\ket{e_-}$ (which is also odd in $\varphi$). At the critical point, the first and second excited states cross, as shown in Fig.~\ref{fig:ferbo_vs_phase}(b). On the opposite side of the crossing (open diamond  in Fig.~\ref{fig:ferbo_vs_phase}(a)), all three states exhibit increased hybridization between Andreev sectors due to the reduced junction transmission. Since the parity in $\varphi$ of $\ket{0}$ and $\ket{1}$ are now opposite, like in a standard fluxonium, the relaxation protection is lost. By the same argument, the forbidden transition is then the $0-2$. This would be easily recognizable in two-tone spectroscopy measurements of the fluxonium spectrum as the transition line intensity vanishes around $\varphi_{\rm ext}=0$, due to the suppressed dispersive shift at this specific point \cite{Ardati2024}. Measuring the vanishing intensity of the $0-1$ transition instead of the $0-2$ one would therefore provide evidence of the FerBo regime in an experiment.

\section{Practical implementation}\label{sec:exp}

The FerBo qubit can be implemented using techniques similar to those for the standard fluxonium, but replacing the Josephson tunnel junction with a superconducting weak link containing a few highly transmissive transport channels. Semiconducting nanowire weak links~\cite{Goffman_2017,Hays2018,Tosi2019} are natural candidates, as their intrinsic capacitance (typically $\lesssim 0.1$ fF) is significantly smaller than that of conventional aluminum tunnel junctions (several fF). In practice the FerBo circuit capacitance would be dominated by the inductor capacitance, implemented either through a Josephson junction array~\cite{Manucharyan2009} or a disordered superconductor~\cite{Rieger2023}. Values of $E_C$ larger than in a fluxonium become accessible, thus facilitating operation in the high-impedance regime. Note that nanowire fluxonium qubits were implemented in Refs.~\cite{pita2020gate,strickland2024gatemonium,pitavidal2025}
, but in the $Z<R_Q$ regime.

For semiconductor-superconductor nanowire implementations, gate voltage tunability introduces an additional noise channel that requires careful filtering~\cite{riechert2025carbon}. Experimental investigation will determine whether this effect can be sufficiently suppressed. A trade-off must also be addressed concerning the second excited state of the device ($\ket{2}$, see Fig.~\ref{fig:ferbo_vs_phase}). In the FerBo regime, this state is coupled to $\ket{1}$ because it has opposite parity and both have a large projection into the same Andreev sector. One should ensure that it cannot be populated, neither thermally nor during pulsed operations of the circuit. Conversely, this issue can be dealt with using post-selection \cite{liu2026,wang2026}.

\section{Conclusions}\label{sec:conclusions}

We have introduced the FerBo qubit, a superconducting circuit that combines a fermionic Andreev degree of freedom with a bosonic LC-circuit mode to achieve simultaneous protection against dielectric losses and flux noise dephasing. The key mechanism behind this robustness is that hybridization between these two types of degrees of freedom creates qubit states with both phase delocalization and disjoint support in Andreev space.

Our analysis identifies a well-defined parameter regime, characterized by high junction transmission and high circuit impedance, where the charge matrix element $|\bra{0}\hat{n}\ket{1}|^2$ and flux dispersion $\partial^2 E_{01}/\partial\phi_{\text{ext}}^2$ are simultaneously suppressed by several orders of magnitude. This protected regime is separated from the unprotected regime by a sharp boundary at $Z/R_Q \approx 2E_C/(\pi\epsilon_r)$, which coincides with a crossing between the first and second excited states.

The protection mechanism relies on a parity transition: in the protected regime, the ground and first excited states have the same parity under phase inversion, suppressing their overlap under charge and phase operators. This same-parity configuration arises from the ground state residing mainly in the $\ket{-}$ Andreev manifold while the first excited state resides mainly in $\ket{+}$.

The FerBo circuit can be realized with existing technology using semiconducting nanowire weak links embedded in high-impedance LC circuits. While practical challenges remain (gate voltage noise, leak to second excited state) the predicted parameter regime for protection is experimentally accessible, making the FerBo qubit a promising candidate for hardware-level quantum error suppression.



\section*{Acknowledgments}

We thank B. Douçot, L. Tosi, M. Houzet, J. Meyer, and S. Deléglise for useful discussions. Support by the ANR grant FerBo and by the Paris Ile-de-France Region in the framework of DIM QuanTiP and by Spanish AEI through grant PID2023-150224NB-I00, as well as through the ``Mar\'{\i}a de Maeztu'' Program for Units of Excellence in R\&D (CEX2018-000805-M), are gratefully acknowledged.

\begin{appendices} 

\section{Description of the FerBo for intermediate transmissions}\label{}

When $\epsilon_r$ or $\delta \Gamma$ is not very small, the ballistic basis is not appropriate to understand the FerBo physics. We show here that the ``Andreev basis'' is more relevant. The term of the Hamiltonian (Eq.~(\ref{eq:potential})) associated with the weak link can be written as $H_{\rm WL}(\hat \varphi)=-\vec{B}(\hat{\varphi}) \cdot \hat{\vec{\sigma}}$, with $\hat{\vec{\sigma}} = (\hat{\sigma}_x, \hat{\sigma}_y, \hat{\sigma}_z)^\intercal$ the vector of Pauli matrices and the effective field $\vec{B}$ defined as:
\begin{equation}
    \vec{B}(\hat{\varphi}) = \left( \Gamma \cos(\hat\varphi/2), -\delta\Gamma \sin(\hat\varphi/2), \epsilon_r\right)^\intercal.
    \label{eq:ferbo-effective-field}
\end{equation}

\begin{figure*}[t]
	\includegraphics[width=0.9\textwidth]{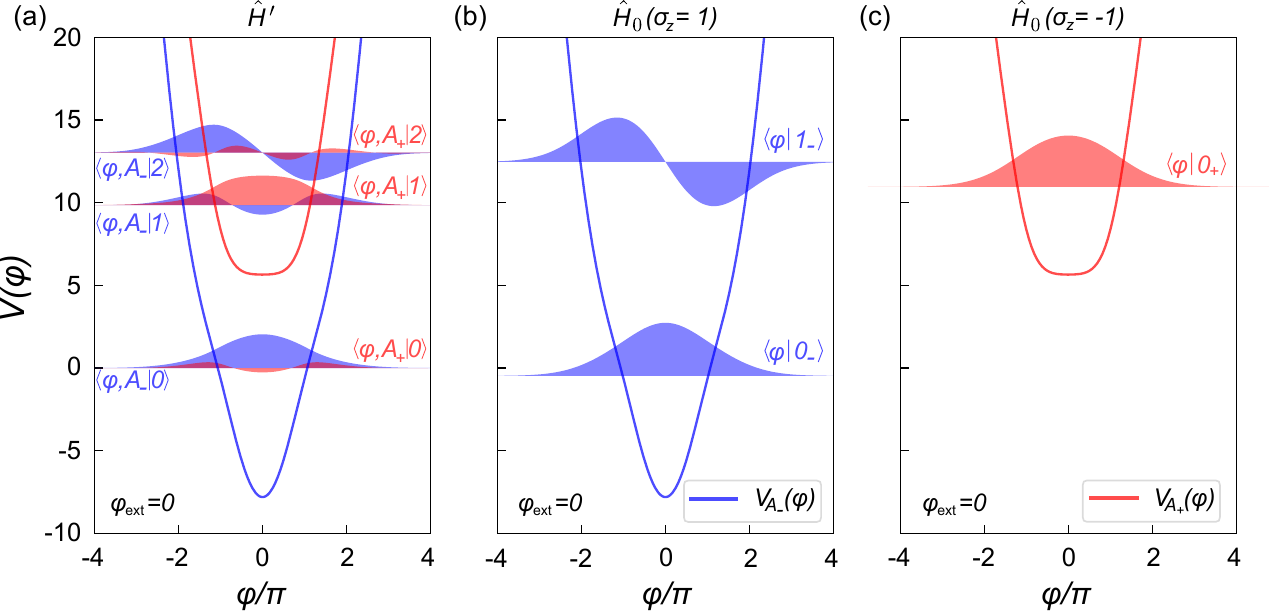}
	\caption{Eigenstates and potential landscapes of the FerBo for an intermediate transmission ($E_C=20$~GHz $E_L=1$~GHz, $\Gamma=5$~GHz, $\delta\Gamma=0$, $\epsilon_r=4.5$~GHz, $\varphi_{\text{ext}}=0$, corresponding to $\tau=0.55$). (a) Full energy spectrum of $\hat{H}'$ showing the probability densities of the wavefunctions of the 3 lowest energy states $\ket{0},$ $\ket{1},$ $\ket{2},$ projected on the Andreev basis: $\braket{\hat \varphi,A_- |i}$, in blue, and $\braket{\hat \varphi,A_+|i}$ in red. The global ground state $\ket{0}$ and first excited state $\ket{1}$ have the same parity (even), protecting them from direct dipole transitions. (b) Potential $V_{A_-}(\varphi)$ for the lower Andreev sector ($\sigma_z=1$) hosting the dominant component of $\ket{0}$. (c) Potential $V_{A_+}(\varphi)$ for the upper Andreev sector ($\sigma_z=-1$) hosting the dominant component of $\ket{1}$.}
	\label{fig:adiabatic-wavefunctions}
\end{figure*}

To analyze the system far from the ballistic limit, we perform a unitary transformation $\hat U(\hat \varphi)$ that rotates the local frame such that the quantization axis ($z$-axis) follows the effective field direction $\vec{B}(\varphi)$. In this basis, which we term the \textit{Andreev basis}, with eigenstates named $\ket{A_+}$ and $\ket{A_-},$ the potential term $-\vec{B}(\varphi) \cdot \hat{\vec{\sigma}}$ is diagonal with elements
\begin{equation}
    \pm B(\hat\varphi) = \pm {\Delta}_{\rm eff} \sqrt{1 - \tau \sin^2(\hat \varphi/2)}.
\end{equation}
Combined with the inductive energy, this defines two distinct potential energy landscapes, or ``Andreev bands'', for the system:
\begin{equation}
    V_{A_\pm}(\varphi) = \frac{1}{2} E_L  \varphi^2 \pm B(\varphi).
\end{equation}
While the potential energy is diagonal in this frame, the kinetic energy term transforms non-trivially due to the phase dependence of the basis change. The transformed Hamiltonian $\hat{H}' = \hat U^\dagger(\hat\varphi) \hat{H}\hat U(\hat\varphi)$ reads:
\begin{equation}
\begin{split}
    \hat{H}' = &4 E_C \left(\hat{n} - \hat{\mathcal{A}}(\hat{\varphi})\right)^2 
    + \dfrac{1}{2}E_L\hat \varphi^2-B(\hat \varphi)\hat \sigma_z.
\end{split}
\end{equation}
Here $\hat{\mathcal{A}}(\hat \varphi) = i \hat U^\dagger(\hat\varphi) \partial_\varphi \hat U(\hat\varphi)$ is the Berry connection \cite{Ivanov1998}. A lengthy calculation gives the following expression of $\hat{\mathcal{A}}(\hat \varphi)$ in the Andreev basis:

\begin{equation}
    \hat{\mathcal A}(\hat\varphi)=\dfrac{\Gamma\delta\Gamma\,}{4\,B(\hat \varphi)\,B_\perp(\hat \varphi)}(\hat \sigma_x+ a\hat \sigma_y+b \hat \sigma_z)\;,
    \label{eq:A_explicit}
\end{equation}
with
\begin{equation}
\begin{split}
&a=- \frac{\epsilon_r}{2B(\hat \varphi)}\left(\frac{\Gamma}{\delta \Gamma}+\frac{\delta \Gamma}{\Gamma}\right)\sin(\hat\varphi),\\
&b=-\frac{\epsilon_r}{B_{\perp}(\hat\varphi)},
\end{split}
\end{equation}
and
\begin{equation}
B_\perp(\hat \varphi)=\sqrt{B(\hat \varphi)^2-\epsilon_r^2}.
\end{equation}

We can separate this Hamiltonian into a diagonal part $\hat{H}_0$, which describes the dynamics within each Andreev band, and an interaction part $\hat{H}_{\text{int}}$ that couples the two sectors:
\begin{equation}
\begin{split}
    &\hat{H}'= \hat{H}_0+\hat{H}_{\text{int}},\\
    &\hat{H}_0=4 E_C \hat{n}^2 + \frac{1}{2} E_L \hat{\varphi}^2 - B(\hat \varphi) \hat{\sigma}_z + 4 E_C \hat{\mathcal{A}}^2(\hat \varphi)\\
    &\hat{H}_{\text{int}}=-4 E_C \{\hat{n}, \hat{\mathcal{A}}(\hat \varphi)\}.
\end{split}
\end{equation}
When $\hat{H}_{\text{int}}$ is small, the eigenstates of the system correspond to those of the independent potentials $V_{A_{-,+}}$ corresponding to $\sigma_z=\mp1$ in $H_0$. If the energy of the first excited state in $V_{A_-}$ is larger than the ground state energy in $V_{A_+}$, the ground and first excited state of $H'$ have even parity and are not coupled by operators like $\hat \varphi$ or $\hat n$.

In the symmetric junction limit $\delta \Gamma =0,$ $\hat{\mathcal{A}}(\hat \varphi)\propto \hat \sigma_y$ so that the correcting term $4 E_C \hat{\mathcal{A}}^2(\hat \varphi)$ in $\hat{H}_0$ is proportional to identity and acts as a state-independent ``diamagnetic'' shift to the potentials. Conversely, $\hat{H}_{\text{int}}$ represents a ``paramagnetic" contribution that couples the Andreev bands. In this limit, $\hat{\mathcal{A}}(\varphi)$ is an odd function of $\varphi$, so that $\{\hat{n}, \hat{\mathcal{A}}(\hat \varphi)\}$ is even and only couples states with the same parity. As a consequence, relaxation only occurs by terms of order $\delta \Gamma/\Gamma.$

In contrast with the high transmission limit, for moderate transmission, since $\hat{H}_0$ is diagonal in the spin basis, it effectively decouples the system into the two independent Andreev sectors. We can thus analyze the dynamics separately for the lower branch ($A_-$) and the upper branch ($A_+$), treating them as distinct 1D problems with potentials $V_{A_-}(\varphi)$ and $V_{A_+}(\varphi)$, respectively (corrected by the diamagnetic term). This separation is visualized in Fig.~\ref{fig:adiabatic-wavefunctions}, where the central and right panels display the independent eigensolutions for each branch. States $\ket{0}$ and $\ket{1}$ have the same parity because they essentially arise from the ground states of the two potentials.

\section{Extended model}
\label{app:tightbinding}

A more microscopic description of the weak link can be obtained by discretizing the Bogoliubov-de Gennes equations for a SNS junction, leading to the following Hamiltonian 

\begin{align}
    H_{TB} = &\sum_{i\sigma} (\epsilon-\mu_i)c_{i\sigma}^\dagger c_{i\sigma} +
    \left( t_{i+1,i} c_{i+1\sigma}^\dagger c_{i\sigma} + h.c. \right) \nonumber \\
    + &\sum_{i}\left( \Delta_i c_{i\downarrow} c_{i\uparrow} + h.c. \right),
\end{align}

\noindent
where $c_{i\sigma}^\dagger$ creates an electron with spin $\sigma$ in site $i$ and the parameters follow from the discretization of the system, which we associate to a conceivable semiconducting-superconducting nanowire junction. The hopping magnitude is in general $t = -\hbar^2/(2m^*a^2)$, with $a=L/N_{\textrm{sites}}$ being the lattice constant and $L$ the length; we take the effective mass of InAs $m^*\approx0.023m_e$ and an extended junction with $L_{N,S}=0.7,1.0$~\SI{}{\micro\meter} ($a{=}50$~nm). At the $NS_{L,R}$ interfaces we allow a different hopping with magnitude $\alpha_{L,R}t$ in order to model a reduced transmission. The onsite potential is $\epsilon=-2t$, from the kinetic term; we set the chemical potential in the superconducting leads to be barely populating the bottom of the band \cite{metzger_thesis2022}, $\mu_{i\in S} = 3\Delta_0$ ($\Delta_{i\in S}=\Delta_0\approx 48$ GHz), and consider the normal region to be tunable by a nearby gate, $\mu_{i\in N} = \mu_N$. The pairing is non-zero only in the leads. 

\begin{figure}[t]
    \centering
    \includegraphics[width=\linewidth]{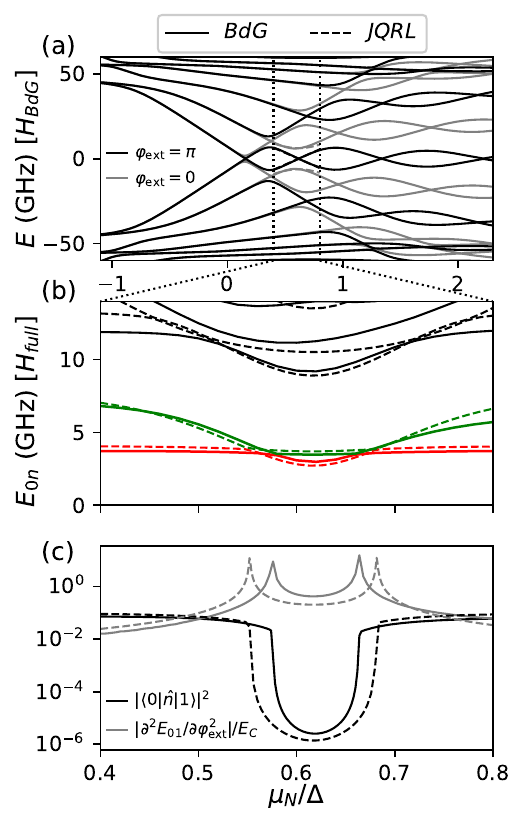}
    \caption{
        Extended Bogoliubov-de Gennes model for a 1D SNS junction as a function of the chemical potential of the normal region. In dashed, results of the JQRL model reproducing a reduced range close to a resonance. (a) Bogoliubov-de Gennes energy spectrum; multiple ABS enter the junction from the pinch-off ($\mu_N{\ll} 0$). (b) Energy transitions of the corresponding circuit with $E_L{=}0.2$GHz, $E_C{=}10$GHz, displaying the FerBo region near the resonance (cf. Fig.~\ref{fig:ferbo_vs_phase}c). (c) Charge matrix element and second order flux dephasing susceptibility (cf. Fig. \ref{fig:spectrum_vs_flux_diff_Z}). Parameters for the BdG and JQRL models are given in the text.   
        }
        \label{fig:TB}
\end{figure}

In Fig. \ref{fig:TB}a we show the Bogoliubov-de Gennes (BdG) energy spectrum as a function of $\mu_N$. In the pinch-off (left), the band in N is not populated and we only see states from the leads at $|E|>\Delta_0$; when $\mu_N$ is increased, ABS detach from the quasi-continuum and lower in energy. Because of the finite $L_N$, several ABS can inhabit the junction at subgap energies, potentially increasing the number of fermionic states that could participate in the FerBo wavefunctions. The anticrossings for $\varphi_{\rm ext}{=}\pi$ at $E=0$ are very small due to a choice of parameters that implements a highly transmissive junction (but still with a finite asymmetry $\alpha_{L,R}{=}1{\pm}0.005$), and we choose to focus on the second one ($\mu_N^0/\Delta_0{=}0.617$) so that there is a finite phase dispersion [hinted by the values for $\varphi_e{=}0$ (gray)]. Around that point we use the reduced JQRL to mimic the low energy part of the BdG spectrum (dashed, analogously for the other panels).

In the BdG semiconducting picture, the ground state corresponds to all negative-energy states filled. Thus, in contrast to the JQRL model, its phase dispersion is not fixed by a single ABS. We could supplement it with a scalar $V(\varphi) = \sum_{n\leq{-2}} E_n(\varphi)$, negative $n$ referring to negative energy levels, to account for that remaining dispersion \cite{Keselman2019}, which can be quite relevant in long junctions \cite{Fatemi2025}. However, in our particular situation the results are similar and for simplicity we just use $\epsilon=l_a(\mu_N{-}\mu_N^0)/\Delta_0$, with the lever arm $l_a=30$~GHz, and $\Gamma_{L,R}=3.1\pm0.015$~GHz. 

In the full circuit with $Z\neq0$, the phase $\hat{\varphi}$ is a quantum variable, so the states are of the kind $\ket{\mathcal{E},\varphi}$, where $\mathcal{E}$ refers to the fermionic sector. In the JQRL model $\ket{\mathcal{E}} \in \{ \ket{0}, \ket{\uparrow\downarrow} \}$ but in the extended model the number of possible many-body states increases exponentially with the number of sites, $2^{2N_{\textrm{sites}}-1}$ (even parity). For this reason it is convenient to rotate to the (many-body) Andreev basis, which provides a set of potentials of increasing energies that can be truncated. We use a discretized phase where $\hat{n} {=} {-}i\partial_\varphi$ produces terms of the kind

\begin{align}
    \sum_\mathcal{E} \hspace{-0.4mm} \ket{\mathcal{E},\varphi_{i{+}1}} \hspace{-0.7mm}\bra{\mathcal{E},\varphi_i} {=} \hspace{-1mm}
    \sum_{A,A'} \hspace{-1mm} \braket{A,\varphi_{i{+}1}|A',\varphi_i} \ket{A,\varphi_{i{+}1}} \hspace{-0.7mm}\bra{A',\varphi_i}, \nonumber
\end{align}

\noindent
where $\ket{A,\varphi_i}$ is the many-body state that diagonalizes $H_{TB}$ at a given phase, i.e. $\ket{A(\varphi_i),\varphi_i}$. For instance, the ground state is $\ket{A{=}GS(\varphi_i)}=\prod_{n\leq-1} \gamma_n^\dagger(\varphi_i) \ket{V}$, where $\ket{V}$ is the vacuum of the quasiparticles ($\gamma_n\ket{V}{=}0\, \forall n$), which does not depend on $\varphi$ \cite{Datta1999}, and $\gamma_n = \Phi_n^\dagger \hat{\Psi}$, with $H_{TB}=\hat{\Psi}^\dagger H_{BdG}\hat{\Psi}$, $\hat{\Psi} = (c_{i\uparrow}, c_{i\downarrow}^\dagger)^T$. The products $\braket{A,\varphi_{i+1}|A',\varphi_i}$ can be written in terms of the quasiparticle wavefunctions $\Phi_n$. Higher Andreev configurations are created from the ground state, e.g. $\ket{e_1} = \gamma_{-1}\gamma_1^\dagger\ket{GS}$. We consider the subspace of $\ket{GS}$ plus the states with two excited quasiparticles across a truncated set of ABS of $N_{\textrm{trunc}}=2$ positive and negative spin-degenerate manifolds. We suppose a homogeneous field along the normal region so we write a phase drop [$t_{i+1,i}{=}e^{i(-\varphi+\varphi_e)/(N_\textrm{sites}^N+1)}t$ for the hoppings within $N$].

In panel (b) we plot the transitions of the full circuit in a zoomed region around $\mu_N\approx\mu_N^0$ and recover the same switch between regimes as in Fig. \ref{fig:ferbo_vs_phase}(c) (note the logscale and the bare energies) when $\mu_N$ moves away from the resonance.
The JQRL model reasonably describes the full system for energy scales within the fermionic potentials from $\ket{GS}$ and $\ket{e}$. Higher potentials, corresponding to the \textit{mixed} excitations where one quasiparticle populates the first (positive) ABS and another populates the second \cite{metzger_thesis2022}, would require models with more than one electronic site.

In panel (c) we plot the charge matrix element and the $E_{01}$ transition curvature over $\varphi_{\rm ext}$, also recovering the features discussed in the main text (Fig. \ref{fig:spectrum_vs_flux_diff_Z}).
Within the JQRL model the circuit states have a well defined symmetry over $\varphi$ when $\comm{H}{\Pi_\varphi}=0$, ($\Pi_\varphi$ inverts the phase, $\Pi_\varphi\hat{\varphi}\Pi_\varphi^\dagger = -\hat{\varphi}$), which requires $\delta\Gamma=0$ at $\varphi_e=0$. In the extended model we can combine phase and spatial symmetry $\Pi=\Pi_\varphi\Pi_x$ ($\Pi_x$ flips the sites over the center of the device; note that spatial symmetry in the JQRL can not be defined independently of $\Pi_\varphi$). When $\tau_L{=}\tau_R$ we have $\comm{H}{\Pi}=0$ and find $\bra{0}\hat{n}\ket{1}{=}0$ in the FerBo region. In summary, the JQRL model captures the lowest energy excitations of microscopic weak link models in the high-transparency limit.

\section{Parameter Dependence of charge relaxation susceptibility $|\langle0|\hat{n}\ket{1}|^2$ }
\label{sec:Gamma&deltaGamma}
\begin{figure*}[t]
    \centering
    \includegraphics[width=\linewidth]{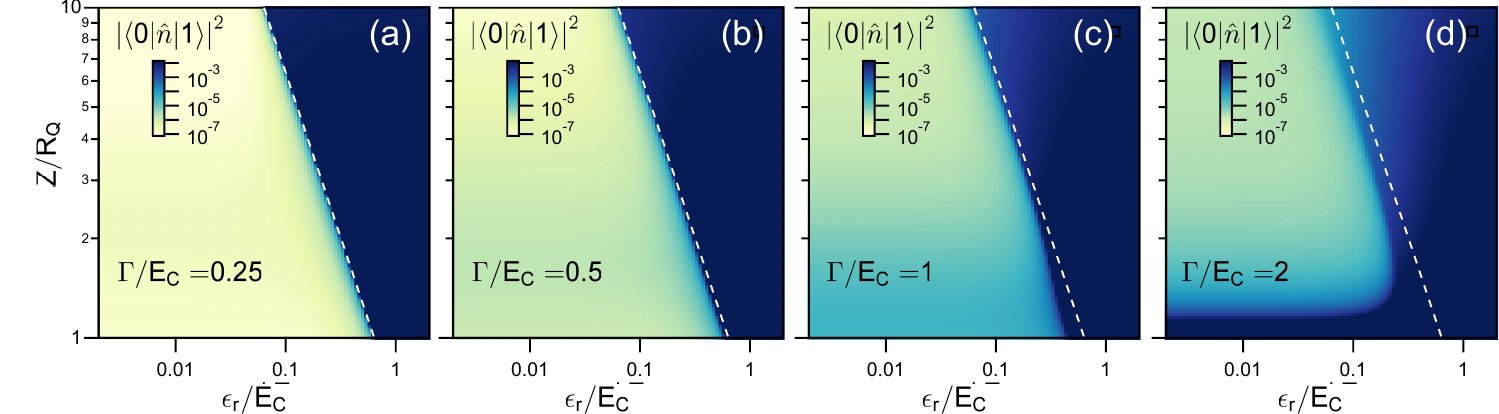}
    \caption{Color map (log scale) of charge relaxation susceptibility $|\bra{0}\hat{n}\ket{1}|^2$ at $\varphi_{ext}=0$ as a function of level position $\epsilon_r/E_C$ and of the reduced impedance $Z/R_Q$ for different $\Gamma/E_C$ ratios: (a) $\Gamma/E_C=0.25$; (b) $\Gamma/E_C=0.5$; (c) $\Gamma/E_C=1$, and (d) $\Gamma/E_C=2$. The white dashed line indicates the approximate boundary $Z/R_Q = 2E_C/(\pi\epsilon_r)$ separating the protected (light-colored, low susceptibility) and unprotected (dark, high susceptibility) regions. The asymmetry is fixed at $\delta\Gamma/\Gamma = 0.01$.
        }
        \label{fig:n_op_ratiosG-Ec}
\end{figure*}

Figure~\ref{fig:n_op_ratiosG-Ec} presents calculations of $|\bra{0}\hat{n}\ket{1}|^2$ for $E_C/\Gamma$ ratios ranging from 0.25 to 2 (0.75 was presented in the main text). The figure demonstrates the generality of the results: a light-colored region indicating low susceptibility appears below a boundary line (white dashed) close to $Z/R_Q=2E_C/(\pi\epsilon_r)$. Note that in the protected region, charge susceptibility continuously increases with this ratio. Furthermore, the actual separation between protected (light-colored) and unprotected (dark) regions deviates from the white dashed line. 

\begin{figure}[t]
    \centering
    \includegraphics[width=\linewidth]{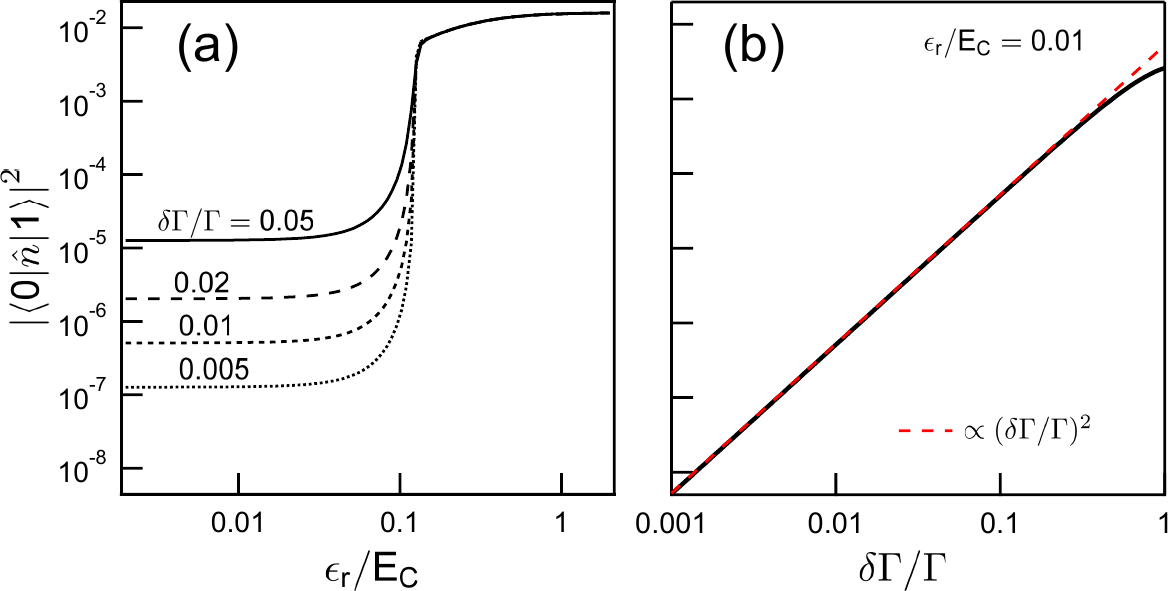}
    \caption{(a) Charge relaxation susceptibility  $|\bra{0}\hat{n}\ket{1}|^2$ at $\varphi_{ext}=0$ as a function of level position $\epsilon_r/E_C$ and of the reduced impedance $Z/R_Q=5$, for different values of $\delta\Gamma/\Gamma$. (b) $|\bra{0}\hat{n}\ket{1}|^2$ at $\varphi_{ext}=0$ as a function of $\delta\Gamma/\Gamma$ for $\epsilon_r/E_C=0.01$ and $Z/R_Q=5$.
        }
        \label{fig:n_op_deltaG}
\end{figure}

Figure~\ref{fig:n_op_deltaG}(a) depicts the dependence of $|\bra{0}\hat{n}\ket{1}|^2$ as a function of level position $\epsilon_r/E_C$ for different values of $\delta\Gamma/\Gamma$. The curves differ in the protected region and merge in the unprotected one. In the protected region, charge susceptibility increases as $(\delta\Gamma/\Gamma)^2$ (Fig.~\ref{fig:n_op_deltaG}(b)) showing that it is critical to keep this ratio as small as possible to benefit from charge relaxation protection.

\end{appendices}

\bibliographystyle{apsrev4-2} 
\bibliography{references} 

\end{document}